\documentclass[12pt]{article}
\usepackage{amssymb}
\input epsf
\makeatletter
\def\numberbysection{\@addtoreset{equation}{section}
 	\def\theequation{\thesection.\arabic{equation}}}
\makeatother

\numberbysection

\newcommand{\be}{\begin{eqnarray}}
\newcommand{\ee}{\end{eqnarray}}
\newcommand{\non}{\nonumber}
\newcommand{\tr}{\mathop{\rm tr}\nolimits}
\newcommand{\id}{\mathbb{I}}

\newcommand{\LL}{\ensuremath{\mathsf{L}}}
\newcommand{\R}{\ensuremath{\mathsf{R}}}
\newcommand{\Z}{\ensuremath{\mathsf{Z}}}

\begin{document}

\begin{titlepage}
\strut\hfill UMTG--226
\vspace{.5in}
\begin{center}

\LARGE Consistent superconformal boundary states\\[1.0in]
\large Rafael I. Nepomechie\\[0.8in]
\large Physics Department, P.O. Box 248046, University of Miami\\[0.2in]  
\large Coral Gables, FL 33124 USA\\

\end{center}

\vspace{.5in}

\begin{abstract}
We propose a supersymmetric generalization of Cardy's equation for
consistent $N=1$ superconformal boundary states.  We solve this
equation for the superconformal minimal models ${\cal SM}(p/p+2)$ with
$p$ odd, and thereby provide a classification of the possible
superconformal boundary conditions .  In addition to the Neveu-Schwarz
($NS$) and Ramond ($R$) boundary states, there are $\widetilde{NS}$
states.  The $NS$ and $\widetilde{NS}$ boundary states are related by
a $Z_{2}$ ``spin-reversal'' transformation.  We treat the tricritical
Ising model as an example, and in an appendix we discuss the
(non-superconformal) case of the Ising model.
\end{abstract}

\end{titlepage}

\setcounter{footnote}{0}

\section{Introduction}\label{sec:intro}

Two fundamental developments of two-dimensional conformal field theory
(CFT) \cite{BPZ, CFT} have been the incorporation of supersymmetry
\cite{SUSY} and the extension to manifolds with boundary \cite{Ca1}.
The concept of conformal boundary state \cite{Ca2} is of
central importance in the formulation of boundary CFT.
Hence, in string theory \cite{Sc, GSW}, boundary states also figure
prominently \cite{string}.  (For further references and recent reviews
of the boundary state formalism for describing D-branes, see e.g.
\cite{Dbranes}.)
The non-supersymmetric (Virasoro algebra) case is well understood
\cite{Ca2}: at ``tree'' level, conformal invariance implies the
constraint
\be
\left(L_{n} - \bar L_{-n} \right) |\alpha \rangle = 0
\label{treeconstraint}
\ee
on the boundary state $|\alpha \rangle $.  This equation has a vector
space of solutions which is spanned by the so-called Ishibashi states
\cite{Is}.  For the conformal minimal models, there is an Ishibashi
state $|j\rangle \rangle$ corresponding to each chiral primary
field $\Phi_{j}(z)$ (or Virasoro highest weight representation with
highest weight $j$),  
\be
| j \rangle\rangle = \sum_{N} | j \,; N \rangle 
\otimes U \overline{| j \,; N \rangle} \,,
\label{Ishibashi}
\ee
where $U$ is an antiunitary operator satisfying 
$U^{\dagger} \bar L_{n} U = \bar L_{n}$, and $| j \,; N \rangle $ is 
an orthonormal basis of the representation.

\begin{figure}[htb]
	\centering
	\epsfxsize=0.4\textwidth\epsfbox{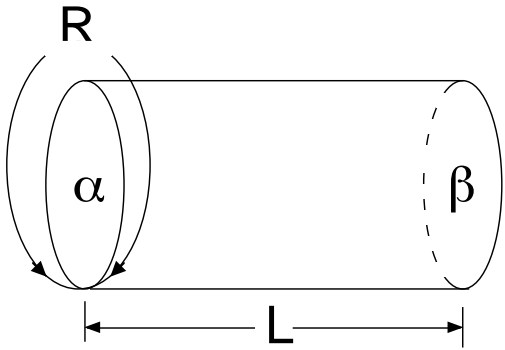}
	\caption[xxx]{\parbox[t]{.6\textwidth}{
	Cylinder of length $\LL$ and circumference $\R$.}
	}
	\label{figcylinder}
\end{figure}

There is a further consistency constraint
\be
\tr e^{-\R H_{\alpha \beta}^{open}} = 
\langle \alpha | e^{-\LL H^{closed}} |\beta \rangle \,,
\label{cylinderconstraint}
\ee 
which arises for the model on a flat cylinder of length $\LL$ and 
circumference $\R$, as represented in Figure \ref{figcylinder}.
Here $H_{\alpha \beta}^{open} = {\pi\over \LL}(L_{0} - {c\over 24})$ is 
the Hamiltonian in the ``open'' channel, with spatial boundary 
conditions denoted by $\alpha$ and $\beta$; and $H^{closed}= 
{2\pi\over \R}\left( L_{0} + \bar L_{0} - {c\over 12} \right)$ is the 
Hamiltonian in the ``closed'' channel.  In the string
literature, a similar constraint (with integrations with respect to 
the corresponding moduli) is known as ``world-sheet duality'' or
``open/closed string duality''.  The
LHS of Eq.  (\ref{cylinderconstraint}) can be expressed as
\be
\tr e^{-\R H_{\alpha \beta}^{open}} = \sum_{i} N_{\alpha \beta}^{i} 
\chi_{i}(q) \,,
\label{lhs0}
\ee
where the Virasoro characters $\chi_{i}(q)$ are defined as
\be
\chi_{i}(q) = \tr_{i} q^{L_{0} - {c\over 24}} \,,
\label{characters}
\ee
and $q=e^{- \pi \R/\LL}$. Under the modular transformation $S$, the 
characters transform according to
\be
\chi_{i}(q) = \sum_{j} S_{i j} \chi_{j}(\tilde q) \,,
\label{modtransf}
\ee
where $\tilde q= e^{- 4 \pi \LL/\R}$. Thus,
\be
\tr e^{-\R H_{\alpha \beta}^{open}} = \sum_{i\,, j} N_{\alpha \beta}^{i} 
 S_{i j} \chi_{j}(\tilde q)\,.
\label{lhs}
\ee
Expressing the RHS of Eq. (\ref{cylinderconstraint}) in the Ishibashi basis, 
one obtains
\be
\langle \alpha | e^{-\LL H^{closed}} |\beta \rangle = \sum_{j}
\langle \alpha | j \rangle \rangle \langle \langle j | \beta \rangle 
\chi_{j}(\tilde q) \,,
\label{rhs}
\ee 
assuming that each representation $j$ appears once in the spectrum 
of $H^{closed}$. Comparing Eqs. (\ref{lhs}) and (\ref{rhs}), one arrives at 
the Cardy equation
\be
\sum_{i} N_{\alpha \beta}^{i} S_{i j} = 
\langle \alpha | j \rangle \rangle \langle \langle j | \beta \rangle 
\,.
\label{cardyeqn}
\ee
Cardy solved this equation for the consistent boundary states
\be
|{\bf k} \rangle = \sum_{j} {S_{k j}\over \sqrt{S_{0 j}}}
| j \rangle \rangle \,.
\label{cardystate}
\ee
Moreover, with the help of the Verlinde formula \cite{Ve}, Cardy
identified $N_{{\bf k} {\bf l}}^{i}$ as the fusion rule coefficients
for $\Phi_{k} \times \Phi_{l} \rightarrow \Phi_{i}$.
The important result (\ref{cardystate}) provides a classification of
the possible conformal boundary conditions for the minimal models, and
gives explicit values for the corresponding $g$-factors \cite{AL},
\be
g_{{\bf k}} = \langle \langle 0 | {\bf k} \rangle =
{S_{k 0}\over \sqrt{S_{0 0}}} \,.
\label{Virgfactor}
\ee 
Renormalization-group (RG) flows between the various conformal
boundary conditions have been investigated in integrable boundary
field theories. (See e.g. \cite{GZ} - \cite{DPTW}, and references 
therein.)

The aim of this paper is to generalize the above considerations to the
case of $N=1$ superconformal field theory \cite{SUSY}, which
encompasses many important models, including superstrings.  Some
progress on this problem has been made by Apikyan and Sahakyan in
\cite{AS1}. We have been motivated in part by our effort to 
better-understand RG boundary flows in supersymmetric integrable  
boundary field theories \cite{AR, AN}.

It is evident that Cardy's results cannot be naively carried over to 
the supersymmetric case. Indeed, (\ref{Virgfactor}) would imply that 
the $g$-factor of any Ramond boundary state is zero, since modular $S$ 
matrix elements between Ramond ($R$) and Neveu-Schwarz ($NS$) 
representations generally vanish (see Eq. (\ref{fullS}) below).

Although for the Virasoro algebra case the consistent boundary states
are in one-to-one correspondence with the irreducible representations,
this is no longer true for the superconformal algebra case.  Indeed,
we find that in the latter case there are more such boundary states . 
This can be traced to the fact that under $S$ modular transformation,
$R$ characters do not transform into $NS$ characters, but rather, into
new characters denoted by $\widetilde{NS}$.  The $NS$ and
$\widetilde{NS}$ Cardy states are related by a $Z_{2}$
``spin-reversal'' transformation, as are the ``fixed $+$'' and ``fixed
$-$'' boundary states of the Ising model.

The outline of this article is as follows.  In Section 2, we briefly
review some necessary results about the $N=1$ superconformal algebra,
its representations, and the modular transformation properties of its
characters.  In Section 3, we formulate a supersymmetric
generalization of Cardy's equation, and we find its solutions.  We
also identify certain coefficients which appear in the super Cardy
equations with the fusion rule coefficients of the chiral primary
superconformal fields.  In Section 4 we work out in detail the case of
the tricritical Ising model (TIM).  This example also serves as a
check on our general formalism, since the TIM is also a member of the
conformal minimal series.  In Section 5, we briefly discuss some
implications of our results, and mention several possible further
generalizations.  In an appendix we present an extended discussion of
the case of the Ising model (IM).  Although the IM does not have
superconformal invariance, it does have $NS$ and $R$ sectors, and it
provides valuable insight into how to treat the sectors of
superconformal models.

\section{Superconformal representation theory}\label{sec:review}

In this Section, we first briefly review the superconformal algebra
and its representations \cite{SUSY}.  We then recall how the
characters \cite{GKO} transform under $S$ modular transformations
\cite{MY, Ka}.

The $N=1$ superconformal algebra is defined by the (anti) commutation 
relations
\be
\left[ L_{m} \,, L_{n} \right] &=& (m-n) L_{m+n} + {1\over 12}c  
(m^{3}-m) \delta_{m+n \,, 0} \,, \non \\
\left[ L_{m} \,, G_{r} \right] &=&  ({1\over 2}m - r) G_{m+r} 
\,, \non \\
\left\{ G_{r} \,, G_{s} \right\} &=& 2 L_{r+s} + {1\over 3} c (r^{2} - 
{1\over 4}) \delta_{r+s \,, 0} \,,
\ee
where $r \,, s \in \Z$ for the Ramond ($R$) sector and $r \,, s \in \Z +
{1\over 2}$ for the Neveu-Schwarz ($NS$) sector. The two modings of 
$G_{r}$ are consistent with the $Z_{2}$ symmetry ($L_{n} \rightarrow 
L_{n} \,, G_{r} \rightarrow -G_{r}$) of the algebra. Highest weight
irreducible representations are generated from highest weight states 
$| \Delta \rangle$ satisfying
\be
L_{0} | \Delta \rangle = \Delta | \Delta \rangle \,, \qquad 
L_{n} | \Delta \rangle = G_{r} | \Delta \rangle = 0 \,, 
\quad n > 0 \,, \quad r > 0 \,.
\ee 
For simplicity, we restrict to the superconformal minimal 
models that are unitary ${\cal SM}(p/p+2)$, 
for which the central charge $c$ has the values
\be
c_{p} = {3\over 2}\left( 1 - {8\over p(p+2)} \right)\,, \qquad 
p = 3 \,, 4 \,, \ldots \,, 
\ee
and the highest weights $\Delta$ are given by
\be
\Delta_{(n \,, m)} = {(n (p+2) - m p)^{2} - 4\over 8 p (p+2)} + 
{1\over 32}(1- (-1)^{n + m}) \,,
\label{weights}
\ee
where $1 \le n \le p-1$ and $1 \le m \le p+1$. The $NS$ representations 
have $n-m$ even, and the $R$ representations have $n-m$ odd. Following 
\cite{MY} \footnote{We generally follow the conventions of Matsuo and
Yahikozawa \cite{MY}, with the main exception that our characters
(\ref{supercharacters}) have an extra factor $q^{- {c\over 24}}$.}
we denote by $\Delta_{NS}$ and $\Delta_{R}$ the following independent
sets of $NS$ and $R$ weights, respectively:
\be 
\Delta_{NS} &=& \Big\{ \Delta_{(n \,, m)} | \ 
1 \le m \le n \le p-1 \,, \quad n-m \quad \mbox{even} \Big\} \,, \non \\
\Delta_{R} &=& \Big\{ \Delta_{(n \,, m)} | \
1 \le m \le n - 1 \quad \mbox{for} \quad 1 < n \le (p-1)/2 \,; \non \\
& & \quad 1 \le m \le n + 1 \quad \mbox{for} \quad (p+1)/2 \le  n \le p-1 \,,
\quad n-m \quad \mbox{odd} \Big\} \,.
\ee 

In the $R$ sector, there is a zero mode $G_{0}$ which commutes with $L_{0}$.
Hence, the highest weight states are generally two-fold degenerate, 
$|\Delta \rangle$ and $G_{0} |\Delta \rangle$. These states have 
opposite Fermion parity, since $G_{0}$ anticommutes with the Fermion parity 
operator $(-1)^{F}$. 
Due to the relation $G_{0}^{2} = L_{0} - {c\over 24}$, 
if $\Delta = {c\over 24}$, then $G_{0} |\Delta \rangle$ is a null state and 
decouples, in which case there is a unique highest weight state.

For $p$ even, there exists a $R$ representation $(n \,, m) = ({p\over 2}
\,, {p+2\over 2})$ which has weight $\Delta_{({p\over 2} \,, 
{p+2\over 2})}= {c_{p}\over 24}$, and so the corresponding highest weight state
is unpaired.  For $p$ odd, all the highest weight states in the $R$
sector are paired.

We define the characters \cite{GKO, MY, Ka}
\be
\chi_{i}^{NS}(q) &=& \tr_{i} q^{L_{0} - {c\over 24}} \,,
\qquad
\chi_{i}^{\widetilde{NS}}(q) = \tr_{i} (-1)^{F} q^{L_{0} - {c\over 24}} 
\,, \quad i \in \Delta_{NS} \,, \non \\
\chi_{i}^{R}(q) &=& \tr_{i} q^{L_{0} - {c\over 24}} \,,
\qquad
\chi_{i}^{\widetilde{R}}= \tr_{i} (-1)^{F} q^{L_{0} - {c\over 24}} \,, 
\quad i \in \Delta_{R} \,.
\label{supercharacters}
\ee
From the above remarks, it follows that for $p$ odd, 
$\chi_{i}^{\widetilde{R} } = 0$ for all representations $i$; and for $p$ even,
\be
\chi_{i}^{\widetilde{R} } = \pm \delta_{i \,, \Delta_{({p\over 2} \,, 
{p+2\over 2})}} \,.
\label{SUSYcase}
\ee

The characters transform under the $S$ modular 
transformation according to \cite{MY, Ka} 
\be
\chi_{i}^{NS}(q) &=& \sum_{j \in \Delta_{NS}} S^{[NS \,, NS]}_{i j}
\chi_{j}^{NS}(\tilde q) \,, \non \\
\chi_{i}^{\widetilde{NS} }(q) &=& \sum_{j \in \Delta_{R}} 
S^{[\widetilde{NS} \,, R]}_{i j}
\sqrt{2} \chi_{j}^{R}(\tilde q) \,, \non \\
\sqrt{2} \chi_{i}^{R}(q) &=& \sum_{j \in \Delta_{NS}} 
S^{[R \,, \widetilde{NS}]}_{i j}
\chi_{j}^{\widetilde{NS} }(\tilde q) \,,
\label{supermodtransf}
\ee
where $\tilde q= e^{- 4 \pi \LL/\R}$.  As already mentioned in the
Introduction, the characters $\chi_{i}^{\widetilde{NS} }$ appear when 
the characters $\chi_{i}^{R}$ undergo a modular transformation.
For the superconformal minimal models ${\cal SM}(p/p+2)$ with $p$ odd,
the modular $S$ matrices are given by
\be
S^{[NS \,, NS]}_{(n \,, m)\,, (n' \,, m')} &=& {4\over \sqrt{p(p+2)}}
\sin{\pi n n' \over p} \sin{\pi m m' \over p+2} \,,  \label{NSmodS}  \\
S^{[\widetilde{NS} \,, R]}_{(n \,, m)\,, (n' \,, m')} &=& {4\over \sqrt{p(p+2)}}
(-1)^{(n-m)/2} \sin{\pi n n' \over p} \sin{\pi m m' \over p+2} \,, 
\label{NSRmodS} \\
S^{[R \,, \widetilde{NS}]}_{(n \,, m)\,, (n' \,, m')} &=& {4\over \sqrt{p(p+2)}}
(-1)^{(n'-m')/2} \sin{\pi n n' \over p} \sin{\pi m m' \over p+2} \,. 
\label{RNSmodS}
\ee
These matrices can be arranged into the matrix $S$
\be
S = \left( \begin{array}{ccc}
   S^{[NS \,, NS]} & 0              & 0 \\
   0               & 0              & S^{[\widetilde{NS} \,, R]} \\
   0               & S^{[R \,, \widetilde{NS}]} & 0 
      \end{array} \right) \,,
\label{fullS}
\ee
which is real, symmetric, and satisfies $S^{2}= \id $.  We do not
quote the corresponding expressions for the case $p$ even, which are
somewhat more complicated due to the special representation 
$({p\over 2} \,, {p+2\over 2})$.

\section{Consistent boundary states}\label{sec:boundstates}
                  
The full operator algebra of the $NS$ and $R$ superconformal primary
fields is nonlocal.  We consider here the so-called spin model
\cite{SUSY} which has a local operator algebra.  It is obtained by
projecting on even Fermion parity $(-1)^{F}=1$ in the $NS$ sector, and
either even or odd Fermion parity $(-1)^{F}= \pm 1$ in the $R$ sector.
\footnote{An analysis of consistent boundary states
for the so-called fermionic model, which is obtained by keeping only
the $NS$ sector, is given in \cite{AS2}.}
For definiteness, we treat only the case with even Fermion parity also 
in the $R$ sector.  Also, for simplicity, we restrict to models all of whose 
representations satisfy $\Delta \ne {c\over 24}$; that is, we consider only
the superconformal minimal models ${\cal SM}(p/p+2)$ with $p$ odd. 
Moreover, we again assume that the bulk theory is diagonal, with each 
representation appearing once.

Our goal is to construct for such spin models the complete set of
consistent superconformal boundary states $|\alpha \rangle$, by
solving the various constraints which they must obey.  The restriction
to even Fermion parity implies the constraint
\be
(-1)^{F} |\alpha \rangle = |\alpha \rangle \,,
\label{spinconstraint}
\ee
where here $F$ is the total Fermion number of right and left movers.
Superconformal invariance implies the constraints \cite{Is, AS1}
\be
\left(L_{n} - \bar L_{-n} \right) |\alpha \rangle &=& 0 \,, \non \\
\left(G_{r} + i \gamma \bar G_{-r} \right) |\alpha \rangle &=& 0 \,,
\label{supertreeconstraint}
\ee
where $\gamma$ is either $+1$ or $-1$.  Finally, we impose the further
constraint
\be
\tr_{NS}{1\over 2}(1 + (-1)^{F}) e^{-\R H_{\alpha \beta}^{open}} 
+\tr_{R}{1\over 2}(1 + (-1)^{F}) e^{-\R H_{\alpha \beta}^{open}} 
= \langle \alpha | e^{-\LL H^{closed}} |\beta \rangle \,,
\label{gencylinderconstraint}
\ee
for a spin model on the cylinder in Figure \ref{figcylinder}.  The
projectors ${1\over 2}(1 + (-1)^{F})$ project onto states of even
Fermion parity in the open channel.  The Hamiltonians in the
open and closed channels are (as in the
non-supersymmetric case which was reviewed in the Introduction)
given by $H_{\alpha \beta}^{open} = {\pi\over \LL}(L_{0} - {c\over 24})$ 
and $H^{closed}= 
{2\pi\over \R}\left( L_{0} + \bar L_{0} - {c\over 12} \right)$, 
respectively. 
This constraint is similar to the Ising-model constraint
(\ref{IMcylinderconstraint}), except without the term involving the
projector ${1\over 2}(1 - (-1)^{F})$ in the $NS$ sector.

We first consider the open channel.  We define the
coefficients $n^{i}_{\alpha \beta}$, etc.  by
\be
\tr_{NS} e^{-\R H_{\alpha \beta}^{open}} &=& 
\sum_{i \in \Delta_{NS}} n^{i}_{\alpha \beta} \chi_{i}^{NS}(q)\,, \non \\
\tr_{NS} (-1)^{F} e^{-\R H_{\alpha \beta}^{open}} &=&  
\sum_{i \in \Delta_{NS}} \widetilde{n}^{i}_{\alpha \beta} 
\chi_{i}^{\widetilde{NS} }(q)
\,, \non \\
\tr_{R} e^{-\R H_{\alpha \beta}^{open}} &=&  
\sum_{i \in \Delta_{R}} m^{i}_{\alpha \beta} \chi_{i}^{R}(q) \,, \non \\
\tr_{R} (-1)^{F} e^{-\R H_{\alpha \beta}^{open}} &=&  
\sum_{i \in \Delta_{R}} \widetilde{m}^{i}_{\alpha \beta} 
\chi_{i}^{\widetilde{R} } = 0 \,,
\label{coeffs}
\ee
where $q=e^{- \pi \R/\LL}$, and the various characters are defined in
(\ref{supercharacters}). In the last line, we have made use of our 
restriction to $p$ odd, together with the result (\ref{SUSYcase}).
It follows that 
\be
\mbox{LHS of Eq. (\ref{gencylinderconstraint})} &=& 
{1\over 2}\sum_{i \in \Delta_{NS}} \left( 
n^{i}_{\alpha \beta} \chi_{i}^{NS}(q) +
\widetilde{n}^{ i}_{\alpha \beta} \chi_{i}^{\widetilde{NS} }(q) \right) 
+ {1\over 2}\sum_{i \in \Delta_{R}}
m^{i}_{\alpha \beta} \chi_{i}^{R}(q)
\non \\ 
&=& 
{1\over 2}\sum_{i \in \Delta_{NS}} \left( 
\sum_{j \in \Delta_{NS}} 
n^{i}_{\alpha \beta} S^{[NS \,, NS]}_{i j}\chi_{j}^{NS}(\tilde q) +
\sum_{j \in \Delta_{R}} 
\widetilde{n}^{ i}_{\alpha \beta} S^{[\widetilde{NS} \,, R]}_{i j}\sqrt{2} 
\chi_{j}^{R}(\tilde q) 
\right)\non  \\
&+& {1\over 2}\sum_{i \in \Delta_{R}}
\sum_{j \in \Delta_{NS}} m^{i}_{\alpha \beta} S^{[R \,, \widetilde{NS}]}_{i j}
{1\over \sqrt{2}} \chi_{j}^{\widetilde{NS}}(\tilde q) 
\,, \label{LHS}
\ee
where $\tilde q= e^{- 4 \pi \LL/\R}$.
In passing to the second equality, we have made use of the modular 
transformation properties (\ref{supermodtransf}) of the characters.

Turning now to the closed channel, we recall
\cite{Is, AS1} that corresponding to each irreducible representation
$j$ of the superconformal algebra, one can construct a pair of Ishibashi
states $|j_{\pm} \rangle \rangle $ satisfying
\be
\left(L_{n} - \bar L_{-n} \right) |j_{\pm} \rangle \rangle &=& 0 \,, \non \\
\left(G_{r} \pm i \bar G_{-r} \right) |j_{\pm} \rangle \rangle &=& 0 \,.
\label{superIsh}
\ee
From the explicit expressions for the Ishibashi states, it is easy to
see that the states in the $NS$ sector have even Fermion parity
\be
(-1)^{F} |j_{\pm}^{NS} \rangle \rangle = |j_{\pm}^{NS} \rangle \rangle 
\label{NSpar}
\,,
\ee
where (as in Eq.  (\ref{spinconstraint})) $F$ is the total Fermion
number of right and left movers.  For the $R$ sector, the computation
of Fermion parity is more subtle due to the presence of zero modes
\cite{AS1}.  We assume that, in analogy with the Ising model result
(\ref{RparIsing}),
\be
(-1)^{F} |j_{\pm}^{R} \rangle \rangle = \pm |j_{\pm}^{R} \rangle \rangle 
\,.
\label{Rpar}
\ee

We propose that the set of Ishibashi states  
$\{ |j_{\pm}^{NS} \rangle \rangle \,, |j_{+}^{R} \rangle \rangle \}$ 
constitutes a basis for the boundary states. That is,
\be
|\alpha \rangle = \sum_{j \in \Delta_{NS}} \left( 
|j_{+}^{NS} \rangle \rangle \langle \langle j_{+}^{NS} |\alpha 
\rangle +
|j_{-}^{NS} \rangle \rangle \langle \langle j_{-}^{NS} |\alpha 
\rangle \right) + 
\sum_{j\in \Delta_{R}} 
|j_{+}^{R} \rangle \rangle \langle \langle j_{+}^{R} |\alpha 
 \rangle \,. 
\label{expansion}
\ee
Indeed, Eqs.  (\ref{NSpar}) and (\ref{Rpar}) imply that the constraint
(\ref{spinconstraint}) is already satisfied.  For a given value of $\gamma$,
the constraints (\ref{supertreeconstraint}) can be satisfied by
keeping in the expansion (\ref{expansion}) only the terms involving
$|j_{\gamma} \rangle \rangle$, i.e. setting 
$\langle \langle j_{-\gamma} |\alpha \rangle$ = 0.  Moreover,
the number of basis vectors (twice the number of $NS$ representations 
plus the number of $R$ representations) is the same as the dimension of 
the vector space on which the full modular $S$ matrix (\ref{fullS}) 
acts, which we expect is the number of consistent boundary states.
The expansion (\ref{expansion}) is also motivated by the 
corresponding result (\ref{IMbasis}) for the Ising model.

In this basis, we have
\be
\mbox{RHS of Eq. (\ref{gencylinderconstraint})} &=& 
\sum_{j \in \Delta_{NS}} \Big(
\langle \alpha 
|j_{+}^{NS} \rangle \rangle \langle \langle j_{+}^{NS} |
e^{-\LL H^{closed}} 
|j_{+}^{NS} \rangle \rangle \langle \langle j_{+}^{NS} |
\beta \rangle \non  \\
&+&
\langle \alpha 
|j_{+}^{NS} \rangle \rangle \langle \langle j_{+}^{NS} |
e^{-\LL H^{closed}} 
|j_{-}^{NS} \rangle \rangle \langle \langle j_{-}^{NS} |
\beta \rangle \non  \\
&+&
\langle \alpha 
|j_{-}^{NS} \rangle \rangle \langle \langle j_{-}^{NS} |
e^{-\LL H^{closed}} 
|j_{+}^{NS} \rangle \rangle \langle \langle j_{+}^{NS} |
\beta \rangle \non  \\
&+&
\langle \alpha 
|j_{-}^{NS} \rangle \rangle \langle \langle j_{-}^{NS} |
e^{-\LL H^{closed}} 
|j_{-}^{NS} \rangle \rangle \langle \langle j_{-}^{NS} |
\beta \rangle \Big) \non  \\
&+& \sum_{j \in \Delta_{R}} 
\langle \alpha 
|j_{+}^{R} \rangle \rangle \langle \langle j_{+}^{R} |
e^{-\LL H^{closed}} 
|j_{+}^{R} \rangle \rangle \langle \langle j_{+}^{R} |
\beta \rangle \non  \\
&=& 
\sum_{j \in \Delta_{NS}} \Big[ \left(
\langle \alpha |j_{+}^{NS} \rangle \rangle 
\langle \langle j_{+}^{NS} |\beta \rangle 
+
\langle \alpha |j_{-}^{NS} \rangle \rangle 
\langle \langle j_{-}^{NS} |\beta \rangle 
\right) \chi_{j}^{NS}(\tilde q) \non  \\
&+& \left(
\langle \alpha |j_{-}^{NS} \rangle \rangle 
\langle \langle j_{+}^{NS} |\beta \rangle 
+
\langle \alpha |j_{+}^{NS} \rangle \rangle 
\langle \langle j_{-}^{NS} |\beta \rangle 
\right) \chi_{j}^{\widetilde{NS}}(\tilde q)
\Big] \non  \\
&+& \sum_{j \in \Delta_{R}} 
\langle \alpha |j_{+}^{R} \rangle \rangle 
\langle \langle j_{+}^{R} |\beta \rangle 
\chi_{j}^{R}(\tilde q) \,.
\label{RHS}
\ee
In passing to the second equality, we have used the relations
\be
\langle \langle j_{\pm}^{NS} | e^{-\LL H^{closed}} 
|j_{\pm}^{NS} \rangle \rangle &=& \chi_{j}^{NS}(\tilde q) \,, \non  \\
\langle \langle j_{\mp}^{NS} | e^{-\LL H^{closed}} 
|j_{\pm}^{NS} \rangle \rangle &=& \chi_{j}^{\widetilde{NS}}(\tilde q) \,, \non  \\
\langle \langle j_{\pm}^{R} | e^{-\LL H^{closed}} 
|j_{\pm}^{R} \rangle \rangle &=& \chi_{j}^{R}(\tilde q) \,, \non  \\
\langle \langle j_{\mp}^{R} | e^{-\LL H^{closed}} 
|j_{\pm}^{R} \rangle \rangle &=& \chi_{j}^{\widetilde{R}}(\tilde q) = 0 \,,
\ee
which are analogous to the results (\ref{closedchanresults}) for the
Ising model.

Comparing Eqs. (\ref{LHS}) and (\ref{RHS}), we arrive at the ``super''
Cardy equations (cf. (\ref{cardyeqn}))
\be
{1\over 2}\sum_{i \in \Delta_{NS}} 
n^{i}_{\alpha \beta} S^{[NS \,, NS]}_{i j} &=&
\langle \alpha |j_{+}^{NS} \rangle \rangle 
\langle \langle j_{+}^{NS} |\beta \rangle +
\langle \alpha |j_{-}^{NS} \rangle \rangle 
\langle \langle j_{-}^{NS} |\beta \rangle \,, \non  \\
{1\over \sqrt{2}}\sum_{i \in \Delta_{NS}} 
\widetilde{n}^{i}_{\alpha \beta} S^{[\widetilde{NS} \,, R]}_{i j} &=&
\langle \alpha |j_{+}^{R} \rangle \rangle 
\langle \langle j_{+}^{R} |\beta \rangle \,, \non  \\
{1\over 2\sqrt{2}}\sum_{i \in \Delta_{R}} 
m^{i}_{\alpha \beta} S^{[R \,, \widetilde{NS}]}_{i j} &=&
\langle \alpha |j_{+}^{NS} \rangle \rangle 
\langle \langle j_{-}^{NS} |\beta \rangle +
\langle \alpha |j_{-}^{NS} \rangle \rangle 
\langle \langle j_{+}^{NS} |\beta \rangle \,.
\label{supercardyeqn}
\ee

We now proceed to solve these equations, together with the constraints
(\ref{supertreeconstraint}), for the consistent superconformal
boundary states.  Defining the state $|{\bf 0}^{NS} \rangle$ as the
solution with
$n^{i}_{{\bf 0}^{NS} {\bf 0}^{NS}} = 
\widetilde{n}^{i}_{{\bf 0}^{NS} {\bf 0}^{NS}} = 
\delta^{i}_{0}$, $m^{i}_{{\bf 0}^{NS} {\bf 0}^{NS}} = 0$, we obtain
\be
|{\bf 0}^{NS} \rangle = {1\over \sqrt{2}} \sum_{j \in \Delta_{NS}}
\sqrt{S^{[NS \,, NS]}_{0 j}} |j_{+}^{NS} \rangle \rangle +
{1\over \sqrt[4]{2}} \sum_{j \in \Delta_{R}}
\sqrt{S^{[\widetilde{NS} \,, R]}_{0 j}} |j_{+}^{R} \rangle \rangle 
\,. 
\ee
We then define the states $|{\bf k}^{NS} \rangle$ and 
$|{\bf k}^{\widetilde{NS}} \rangle$ with $k \in \Delta_{NS}$ by 
\be
n^{i}_{{\bf 0}^{NS} {\bf k}^{NS}} &=& 
\widetilde{n}^{i}_{{\bf 0}^{NS} {\bf k}^{NS}} = 
\delta^{i}_{k} \,, \qquad m^{i}_{{\bf 0}^{NS} {\bf k}^{NS}} = 0 \,, \non \\
n^{i}_{{\bf 0}^{NS} {\bf k}^{\widetilde{NS}}} &=& 
-\widetilde{n}^{i}_{{\bf 0}^{NS} {\bf k}^{\widetilde{NS}}} = 
\delta^{i}_{k} \,, \qquad m^{i}_{{\bf 0}^{NS} {\bf k}^{\widetilde{NS}}} = 0 \,,
\ee 
respectively, and we obtain
\be
|{\bf k}^{NS} \rangle &=& {1\over \sqrt{2}} \sum_{j \in \Delta_{NS}}
{S^{[NS \,, NS]}_{k j}\over \sqrt{S^{[NS \,, NS]}_{0 j}}} 
|j_{+}^{NS} \rangle \rangle +
{1\over \sqrt[4]{2}} \sum_{j \in \Delta_{R}}
{S^{[\widetilde{NS} \,, R]}_{k j}\over 
\sqrt{S^{[\widetilde{NS} \,, R]}_{0 j}}} |j_{+}^{R} \rangle \rangle 
\,, \label{NSstate} \\
|{\bf k}^{\widetilde{NS}} \rangle &=& {1\over \sqrt{2}} 
\sum_{j \in \Delta_{NS}}
{S^{[NS \,, NS]}_{k j}\over \sqrt{S^{[NS \,, NS]}_{0 j}}} 
|j_{+}^{NS} \rangle \rangle -
{1\over \sqrt[4]{2}} \sum_{j \in \Delta_{R}}
{S^{[\widetilde{NS} \,, R]}_{k j}\over 
\sqrt{S^{[\widetilde{NS} \,, R]}_{0 j}}} |j_{+}^{R} \rangle \rangle 
\,. 
\label{tildeNSstate}
\ee
Finally, we define the states $|{\bf k}^{R} \rangle$ 
with $k \in \Delta_{R}$ by 
\be
n^{i}_{{\bf 0}^{NS} {\bf k}^{R}} = 
\widetilde{n}^{i}_{{\bf 0}^{NS} {\bf k}^{R}} = 0
\,, \qquad m^{i}_{{\bf 0}^{NS} {\bf k}^{R}} = 2 \delta^{i}_{k} \,, 
\ee 
and we obtain
\be
|{\bf k}^{R} \rangle = 
\sum_{j \in \Delta_{NS}}
{S^{[R \,, \widetilde{NS}]}_{k j}\over \sqrt{S^{[NS \,, NS]}_{0 j}}} 
|j_{-}^{NS} \rangle \rangle 
\,.
\label{Rstate}
\ee
We shall refer to the states (\ref{NSstate}), (\ref{tildeNSstate}) and
(\ref{Rstate}) as the $NS$, $\widetilde{NS}$ and $R$ 
Cardy states, respectively. These states
manifestly satisfy the constraints (\ref{supertreeconstraint}), with
the $R$ states and the $NS$, $\widetilde{NS}$ states having opposite 
signs of $\gamma$. The $NS$ and $\widetilde{NS}$ states differ by the 
$Z_{2}$ ``spin-reversal'' transformation 
$| j^{NS} \rangle \rangle \rightarrow | j^{NS} \rangle \rangle \,, 
\quad | j^{R} \rangle \rangle \rightarrow -| j^{R} \rangle \rangle$, 
just like the ``fixed $+$'' and ``fixed 
$-$'' boundary states of the Ising model (\ref{IMstates}).

The Eqs.  (\ref{supercardyeqn}) and their solutions (\ref{NSstate}),
(\ref{tildeNSstate}), (\ref{Rstate}) are the main results of this
paper. \footnote{In \cite{AS1} a different set of equations is
proposed, which gives the $NS$ states (\ref{NSstate}), but not
the $\widetilde{NS}$ and $R$ states (\ref{tildeNSstate}),
(\ref{Rstate}).}
These solutions provide a classification of the possible
superconformal boundary conditions for the superconformal minimal
models ${\cal SM}(p/p+2)$ with $p$ odd.

The $g$-factor \cite{AL} of a boundary state $|\alpha \rangle$ is
given by
\be
g_{\alpha} = \left( 
\langle \langle 0_{+}^{NS} | + \langle \langle 0_{-}^{NS} | \right) 
|\alpha \rangle \,.
\label{gfactor}
\ee
Hence, the $g$-factors of the Cardy states are
\be
g_{{\bf k}^{NS}} &=& g_{{\bf k}^{\widetilde{NS}}} = 
{1\over \sqrt{2}} 
{S^{[NS \,, NS]}_{k 0}\over \sqrt{S^{[NS \,, NS]}_{0 0}}} \,, 
\label{NSgfacator}  \\
g_{{\bf k}^{R}} &=&  
{S^{[R \,, \widetilde{NS}]}_{k 0}\over \sqrt{S^{[NS \,, NS]}_{0 0}}} \,.
\label{Rgfactor}
\ee 
We see from (\ref{NSgfacator}) that, for a $NS$ state, the naive use
of the modular $S$ matrix (\ref{NSmodS}) in the original Cardy result
(\ref{Virgfactor}) would give a $g$-factor which is a factor
$\sqrt{2}$ too big.  Moreover, the $g$-factor (\ref{Rgfactor}) of a
$R$ state does {\it not} generally vanish.

As in the non-supersymmetric case, the various coefficients
$n^{i}_{\alpha \beta}$, etc.  in Eq.  (\ref{coeffs}) can now be
expressed in terms of modular $S$ matrices and be related to fusion rule
coefficients.  Indeed, by substituting the expression (\ref{NSstate})
for two $NS$ Cardy states $| {\bf k}^{NS} \rangle$ and $| {\bf l}^{NS} \rangle$
back into the super Cardy formula (\ref{supercardyeqn}), we obtain
\be
n^{i}_{{\bf k}^{NS} {\bf l}^{NS}} &=& \sum_{j \in \Delta_{NS}} 
{S^{[NS \,, NS]}_{k j} S^{[NS \,, NS]}_{l j} 
(S^{[NS \,, NS]})^{-1}_{j i} \over S^{[NS \,, NS]}_{0 j}} \,, \non  \\
\widetilde{n}^{i}_{{\bf k}^{NS}{\bf l}^{NS}} 
&=& \sum_{j \in \Delta_{R}} 
{S^{[\widetilde{NS} \,, R]}_{k j} S^{[\widetilde{NS} \,, R]}_{l j} 
(S^{[\widetilde{NS} \,, R]})^{-1}_{j i} 
\over S^{[\widetilde{NS} \,, R]}_{0 j}} \,, \non  \\ 
m^{i}_{{\bf k}^{NS} {\bf l}^{NS}} &=& 0 \,.
\label{NSNSfusion}
\ee
From the work \cite{EH} (see also \cite{AANN}) on a generalized
Verlinde formula, we can identify $n^{i}_{{\bf k}^{NS} {\bf l}^{NS}}$
as the fusion rule coefficient for
$\Phi^{NS}_{k} \times \Phi^{NS}_{l} \rightarrow \Phi^{NS}_{i}$.
Similarly, for two $R$ Cardy states (\ref{Rstate}), we obtain
\be
n^{i}_{{\bf k}^{R} {\bf l}^{R}} &=& 2 \sum_{j \in \Delta_{NS}} 
{S^{[R \,, \widetilde{NS}]}_{k j} S^{[R \,, \widetilde{NS}]}_{l j} 
(S^{[NS \,, NS]})^{-1}_{j i} \over S^{[NS \,, NS]}_{0 j}} \,, \non  \\ 
\widetilde{n}^{i}_{{\bf k}^{R} {\bf l}^{R}} &=& 
m^{i}_{{\bf k}^{R} {\bf l}^{R}} = 0 \,, 
\label{RRfusion}
\ee
and we identify $n^{i}_{{\bf k}^{R} {\bf l}^{R}}$ 
as the fusion rule coefficient for 
$\Phi^{R}_{k} \times \Phi^{R}_{l} \rightarrow \Phi^{NS}_{i}$. Finally, 
for one $NS$ state and one $R$ state, we obtain 
\be
m^{i}_{{\bf k}^{NS} {\bf l}^{R}} &=& 2 \sum_{j \in \Delta_{NS}} 
{S^{[NS \,, NS]}_{k j} S^{[R \,, \widetilde{NS}]}_{l j} 
(S^{[R \,, \widetilde{NS}]})^{-1}_{j i} \over S^{[NS \,, NS]}_{0 j}} 
\,, \non  \\ 
n^{i}_{{\bf k}^{NS} {\bf l}^{R}} &=&
\widetilde{n}^{i}_{{\bf k}^{NS} {\bf l}^{R}} = 0 \,, 
\label{NSRfusion}
\ee
and we identify  $m^{i}_{{\bf k}^{NS} {\bf l}^{R}}$ 
as the fusion rule coefficient for 
$\Phi^{NS}_{k} \times \Phi^{R}_{l} \rightarrow \Phi^{R}_{i}$.  The
results for the coefficients involving $\widetilde{NS}$ states 
(\ref{tildeNSstate}) are very similar to those for the corresponding 
$NS$ states.

\section{Tricritical Ising model}\label{sec:TIM}

As an example of the general formalism presented in the previous
section, we now work out in detail the first nontrivial case: namely,
the superconformal minimal model ${\cal SM}(3/5)$ ($p=3$), which has been
identified \cite{SUSY} as the tricritical Ising model (TIM). This model is 
equivalent to the conformal minimal model ${\cal M}(4/5)$, for which 
the Cardy states are already known \cite{Ca2, Chim}. Hence, this 
example also serves as a valuable check on our general formalism. 

The Kac table for ${\cal SM}(3/5)$, which is obtained using Eq. 
(\ref{weights}), is given in Table \ref{figSM35}.
\begin{table}[htb] 
  \centering
  \begin{tabular}{|c|c|c|c|c|c|c|}\hline
    $\frac{7}{16}$ & $\frac{1}{10}$ & $\frac{3}{80}$ & 0 \\
    \hline
    0 & $\frac{3}{80}$ & $\frac{1}{10}$ & $\frac{7}{16}$ \\
    \hline
   \end{tabular}
  \caption{Kac table for ${\cal SM}(3/5)$}
  \label{figSM35}
\end{table}
The modular $S$ matrix is (\ref{NSmodS}) - (\ref{fullS})
\be
S= \left( \begin{array}{cccccc}
2a &  2b & 0 & 0 & 0 & 0 \\
2b & -2a & 0 & 0 & 0 & 0 \\
0  &   0 & 0 & 0 &\sqrt{2}  c &\sqrt{2} d \\
0  &   0 & 0 & 0 &-\sqrt{2} d &\sqrt{2} c \\
0  &   0 & \sqrt{2}  c &-\sqrt{2} d & 0 & 0 \\
0  &   0 & \sqrt{2}  d &\sqrt{2} c & 0 & 0 
\end{array} \right) \,,
\label{SM35modS}
\ee 
where the rows and columns are labeled by the highest weights
$(0^{NS}\,, {1\over 10}^{NS} \,, 
0^{\widetilde{NS}} \,, {1\over 10}^{\widetilde{NS}} \,, 
{7\over 16}^{R} \,, {3\over 80}^{R})$, and where 
\be
a &=& {1\over 2}\sqrt{{1\over 10}(5-\sqrt{5})} \,, \qquad 
b = {1\over 2}\sqrt{{1\over 10}(5+\sqrt{5})} \,, \non \\
c &=& {1\over 2}\sqrt{{1\over 5}(5-\sqrt{5})} \,, \qquad 
d = {1\over 2}\sqrt{{1\over 5}(5+\sqrt{5})} \,. 
\label{abcd}
\ee
According to our results (\ref{NSstate}), (\ref{tildeNSstate}),
(\ref{Rstate}), there are 6 Cardy states, given by
\be
|{\bf 0}^{NS} \rangle &=& \sqrt{a} |0^{NS}_{+} \rangle \rangle
+ \sqrt{b} |{1\over 10}^{NS}_{+} \rangle \rangle 
+ \sqrt{c} |{7\over 16}^{R}_{+} \rangle \rangle 
+ \sqrt{d} |{3\over 80}^{R}_{+} \rangle \rangle \,, \non \\
|{\bf 0}^{\widetilde{NS}} \rangle &=& \sqrt{a} |0^{NS}_{+} \rangle \rangle
+ \sqrt{b} |{1\over 10}^{NS}_{+} \rangle \rangle 
- \sqrt{c} |{7\over 16}^{R}_{+} \rangle \rangle 
- \sqrt{d} |{3\over 80}^{R}_{+} \rangle \rangle \,, \non \\
|{\bf {1\over 10}}^{NS} \rangle &=& {b\over \sqrt{a}} |0^{NS}_{+} \rangle \rangle
- {a\over \sqrt{b}} |{1\over 10}^{NS}_{+} \rangle \rangle 
- {d\over \sqrt{c}} |{7\over 16}^{R}_{+} \rangle \rangle 
+ {c\over \sqrt{d}} |{3\over 80}^{R}_{+} \rangle \rangle \,, \non \\
|{\bf {1\over 10}}^{\widetilde{NS}} \rangle &=& 
{b\over \sqrt{a}} |0^{NS}_{+} \rangle \rangle
- {a\over \sqrt{b}} |{1\over 10}^{NS}_{+} \rangle \rangle 
+ {d\over \sqrt{c}} |{7\over 16}^{R}_{+} \rangle \rangle 
- {c\over \sqrt{d}} |{3\over 80}^{R}_{+} \rangle \rangle \,, \non \\
|{\bf {7\over 16}}^{R} \rangle &=& {c\over \sqrt{a}} |0^{NS}_{-} \rangle \rangle
- {d\over \sqrt{b}} |{1\over 10}^{NS}_{-} \rangle \rangle \,, \non \\
|{\bf {3\over 80}}^{R} \rangle &=& {d\over \sqrt{a}} |0^{NS}_{-} \rangle \rangle
+ {c\over \sqrt{b}} |{1\over 10}^{NS}_{-} \rangle \rangle \,.
\label{SM35states}
\ee 
Using (\ref{gfactor}), we obtain the $g$ factors
\be
g_{{\bf 0}^{NS}} &=& g_{{\bf 0}^{\widetilde{NS}}}= \sqrt{a} \,, \qquad 
g_{{\bf {1\over 10}}^{NS}} = g_{{\bf {1\over 10}}^{\widetilde{NS}}}
= {b\over \sqrt{a}} \,, \non  \\
g_{{\bf {7\over 16}}^{R} } &=& {c\over \sqrt{a}} \,, \qquad
g_{{\bf {3\over 80}}^{R} } = {d\over \sqrt{a}} 
\,. 
\ee

Let us compare these results with those \cite{Ca2, Chim} obtained from 
the ${\cal M}(4/5)$ description.  The ${\cal M}(4/5)$ Kac table is 
given in Table \ref{figM45}.
\begin{table}[htb] 
  \centering
  \begin{tabular}{|c|c|c|c|}\hline
    $\frac{3}{2}$ & $\frac{3}{5}$ & $\frac{1}{10}$ & 0 \\
    \hline
    $\frac{7}{16}$ & $\frac{3}{80}$ & $\frac{3}{80}$ & $\frac{7}{16}$ \\
    \hline
    0 & $\frac{1}{10}$ & $\frac{3}{5}$ & $\frac{3}{2}$ \\
    \hline
   \end{tabular}
  \caption{Kac table for ${\cal M}(4/5)$}
  \label{figM45}
\end{table}

The modular $S$ matrix is
\be
S= \left( \begin{array}{cccccc}
a &  b & b & a & c & d \\
b & -a & -a & b & -d & c \\
b & -a & -a & b & d & -c \\
a &  b & b & a & -c & -d \\
c & -d & d & -c & 0 & 0 \\
d & c & -c & -d & 0 & 0 \\
\end{array} \right) \,,
\label{M45modS}
\ee 
where the rows and columns are labeled by the highest weights
$(0\,, {1\over 10} \,, {3\over 5} \,,  {3\over 2} \,, {7\over 
16} \,, {3\over 80})$, and $a$-$d$ are given by (\ref{abcd}).
As follows from (\ref{cardystate}), the Cardy states are given by 
\footnote{Our notation is related to Chim's \cite{Chim}
$C=\sqrt{{\sin(\pi/5)\over \sqrt{5}}}$, 
$\eta=\sqrt{{\sin(2\pi/5)\over \sin(\pi/5)}}$ by
\be
a = C^{2} \,,\quad  b= C^{2} \eta^{2} \,, \quad 
c = C^{2} \sqrt{2} \,, \quad d = C^{2}\eta^{2} \sqrt{2} \,. \non 
\ee}
\be
|{\bf 0} \rangle &=& \sqrt{a} |0 \rangle \rangle
+ \sqrt{b} |{1\over 10} \rangle \rangle
+ \sqrt{b} |{3\over 5} \rangle \rangle
+ \sqrt{a} |{3\over 2} \rangle \rangle
+ \sqrt{c} |{7\over 16} \rangle \rangle 
+ \sqrt{d} |{3\over 80} \rangle \rangle \,, \non \\
|{\bf {3\over 2}} \rangle &=& \sqrt{a} |0 \rangle \rangle
+ \sqrt{b} |{1\over 10} \rangle \rangle
+ \sqrt{b} |{3\over 5} \rangle \rangle
+ \sqrt{a} |{3\over 2} \rangle \rangle
- \sqrt{c} |{7\over 16} \rangle \rangle 
- \sqrt{d} |{3\over 80} \rangle \rangle \,, \non \\
|{\bf {1\over 10}} \rangle &=& {b\over \sqrt{a}} |0 \rangle \rangle
- {a\over \sqrt{b}} |{1\over 10} \rangle \rangle 
- {a\over \sqrt{b}} |{3\over 5} \rangle \rangle 
+ {b\over \sqrt{a}} |{3\over 2} \rangle \rangle
- {d\over \sqrt{c}} |{7\over 16} \rangle \rangle 
+ {c\over \sqrt{d}} |{3\over 80} \rangle \rangle \,, \non \\
|{\bf {3\over 5}} \rangle &=& {b\over \sqrt{a}} |0 \rangle \rangle
- {a\over \sqrt{b}} |{1\over 10} \rangle \rangle 
- {a\over \sqrt{b}} |{3\over 5} \rangle \rangle 
+ {b\over \sqrt{a}} |{3\over 2} \rangle \rangle
+ {d\over \sqrt{c}} |{7\over 16} \rangle \rangle 
- {c\over \sqrt{d}} |{3\over 80} \rangle \rangle \,, \non \\
|{\bf {7\over 16}} \rangle &=& {c\over \sqrt{a}} |0 \rangle \rangle
- {d\over \sqrt{b}} |{1\over 10} \rangle \rangle 
+ {d\over \sqrt{b}} |{3\over 5} \rangle \rangle 
- {c\over \sqrt{a}} |{3\over 2} \rangle \rangle
\,, \non  \\
|{\bf {3\over 80}} \rangle &=& {d\over \sqrt{a}} |0 \rangle \rangle
+ {c\over \sqrt{b}} |{1\over 10} \rangle \rangle 
- {c\over \sqrt{b}} |{3\over 5} \rangle \rangle 
- {d\over \sqrt{a}} |{3\over 2} \rangle \rangle \,.
\label{M45states}
\ee 

We observe that the two modular $S$ matrices (\ref{SM35modS}) and 
(\ref{M45modS}) are related by a unitary transformation, due to the 
relation of the corresponding characters \cite{MY}
\be
\chi^{NS}_{0}(q) &=& \chi_{0}(q) + \chi_{{3\over 2}}(q) \,, \qquad 
\chi^{\widetilde{NS}}_{0}(q) = \chi_{0}(q) - \chi_{{3\over 2}}(q) \,, \non \\
\chi^{NS}_{{1\over 10}}(q) &=& 
\chi_{{1\over 10}}(q) + \chi_{{3\over 5}}(q)\,, \qquad 
\chi^{\widetilde{NS}}_{{1\over 10}}(q) =
\chi_{{1\over 10}}(q) - \chi_{{3\over 5}}(q) \,,\non  \\
\chi_{{7\over 16}}^{R}(q) &=& \chi_{{7\over 16}}(q) \,, \qquad 
\chi_{{3\over 80}}^{R}(q) = \chi_{{3\over 80}}(q) \,.
\ee
Moreover, the Cardy states (\ref{SM35states}) and (\ref{M45states}) 
can be seen to coincide, upon identifying the Ishibashi states
\be
|0^{NS}_{\pm} \rangle \rangle  &=& 
|0 \rangle \rangle \pm |{3\over 2} \rangle \rangle \,, \non \\ 
|{1\over 10}^{NS}_{\pm} \rangle \rangle  &=& 
|{1\over 10} \rangle \rangle \pm |{3\over 5} \rangle \rangle \,, \non \\ 
|{7\over 16}^{R}_{+} \rangle \rangle  &=& |{7\over 16} \rangle \rangle 
\,, \qquad 
|{3\over 80}^{R}_{+} \rangle \rangle  = |{3\over 80} \rangle \rangle \,.
\ee 
Evidently, whether we use the ${\cal M}(4/5)$ or ${\cal SM}(3/5)$ 
description, the Hamiltonian is the same, and so are the Cardy states 
and corresponding $g$ factors. The two descriptions correspond to two 
equivalent bases.

\section{Discussion}\label{sec:discuss}

We have proposed (\ref{supercardyeqn}) a supersymmetric generalization
of Cardy's equation, and we have solved it for the consistent
superconformal boundary states (\ref{NSstate}), (\ref{tildeNSstate}),
(\ref{Rstate}), thereby classifying the possible superconformal
boundary conditions. In particular, there are $\widetilde{NS}$ 
boundary states in addition to the $NS$ and $R$ states.

Having a better understanding of boundary conditions in boundary
superconformal field theories, one is in a better position to
investigate integrable perturbations of these theories, and treat
problems such as RG boundary flows.

For simplicity, we have restricted here to the unitary superconformal
minimal models ${\cal SM}(p/p+2)$ with $p$ odd.  It should be possible
to extend our analysis to the models with $p$ even, and in fact, to
general (nonunitary) models ${\cal SM}(p/q)$.  Also, a similar
analysis should be possible for $N=2$ superconformal models, which are
important for superstring compactifications with spacetime
supersymmetry \cite{Ge, RS}.

\section*{Acknowledgments}

I thank C. Ahn for his collaboration at an early stage of this 
work, C. Efthimiou for bringing references \cite{AS1, AS2} to my 
attention, and S. Apikyan for helpful correspondence. 
This work was supported in part by the National Science
Foundation under Grant PHY-9870101.

\appendix

\section{Ising model}\label{sec:IM}

Although the critical Ising model (i.e., the conformal minimal model
${\cal M}(3/4)$) does not have superconformal symmetry, it does have
$NS$ and $R$ sectors.  Here we work out explicitly how these sectors
``transform'' between the open and closed channels of the
cylinder.  Because the Ising model is a free-field theory, the
computations are particularly simple.  Nevertheless, this exercise is
useful, since it gives insight into how to treat the sectors of a
superconformal model.  Although the Ising model on a cylinder has
already been studied extensively \cite{Ca2, Ch, LMSS, Ya}, this
particular aspect does not seem to have been emphasized before.

The critical two-dimensional Ising model (IM) is described by a free
Majorana spinor field, whose two components we denote by $\psi(x\,,
y)$ and $\bar \psi(x\,, y)$.  We consider this model on the cylinder
shown in Figure \ref{figcylinder}, with $x \in [0 \,, \LL]$ the
coordinate along the axis, and $y \in [0 \,, \R]$ the coordinate along
the circumference.

\subsection{Open channel}\label{subsec:open}

In the open channel, we regard $x$ as the space coordinate and
$y$ as the time coordinate.  The time coordinate thus has period $\R$,
corresponding to the temperature $T = 1/\R$.  The conformally-invariant
spatial boundary conditions (BC) are \cite{Ch}
\be
\psi(0 \,, y) + a i \bar \psi(0 \,, y) &=& 0 \non \\
\psi(\LL \,, y) - b i \bar \psi(\LL \,, y) &=& 0 \,,
\ee
where $a \,, b = +1$ corresponds to ``fixed'' BC, and 
$a \,, b = -1$ corresponds to ``free'' BC. 

Our first task is to find appropriate mode expansions for the fields
$\psi$ and $\bar \psi$.  To this end, we recall that the overall
relative sign between these fields is a matter of convention.  We can
therefore redefine $\bar \psi(x\,, y)$ such that
\be
\psi(0 \,, y) = i \bar \psi(0 \,, y) \,,
\label{BC1}
\ee
which implies
\be
\psi(\LL \,, y) = -i {b\over a} \bar \psi(\LL \,, y) \,.
\label{BC2}
\ee
Proceeding as in the case of the superstring \cite{Sc}, we extend the 
definition of $x$ to $[-\LL \,, \LL]$ and define the new field
\be
\Psi(x \,, y) = \left\{ \begin{array}{lll}
                  \psi(x\,, y) & \mbox{if} & x \in [0 \,, \LL] \\
                 i\bar \psi(-x\,, y) & \mbox{if} & x \in [-\LL \,, 0]
		 \end{array} \right.
\ee
This definition is consistent by virtue of Eq. (\ref{BC1}). It 
follows that $\Psi(x \,, y)$ obeys the (quasi)periodicity condition
\be
\Psi(\LL \,, y) = - {b\over a} \Psi(-\LL \,, y) \,.
\ee
Thus, $\Psi$ is periodic ($R$) if $a=-b$, and $\Psi$ is antiperiodic ($NS$)
if $a=b$. Note that a given set $(a \,, b)$ of boundary conditions is 
compatible with only one ($R$ or $NS$) sector. The field $\Psi$ has the 
standard mode expansion
\be
\Psi(x \,, y) = {1\over \sqrt{2\LL}} \sum_{k} b_{k}
e^{-i {\pi\over \LL} k (x + i y)} \,, \qquad 
\left\{ b_{k} \,, b_{l} \right\} = \delta_{k+l\,, 0} \,,
\ee
with $k \in \Z$ for $R$, and $k \in \Z + {1\over 2}$ for $NS$. It follows 
that the sought-after mode expansions for $\psi$ and $\bar \psi$ are
\be
\psi(x\,,y) &=& {1\over \sqrt{2\LL}} \sum_{k} b_{k}
e^{-i {\pi\over \LL} k (x + i y)} \,, \non \\
\bar \psi(x\,, y )&=& -{i\over \sqrt{2\LL}} \sum_{k} b_{k}
e^{-i {\pi\over \LL} k (-x + i y)} \,.
\ee
There is only one independent set of modes $\left\{ b_{k} \right\}$ 
in the open channel.

The Hamiltonian $H^{open}$ is
\be
H^{open} = {\pi\over \LL}\left(e_{0} + \sum_{k >0} k b_{-k} b_{k} 
\right) = {\pi\over \LL}\left( L_{0} - {c\over 24} \right) \,,
\ee
with 
\be
e_{0}^{NS} = -{1\over 48}, \qquad e_{0}^{R} = {1\over 24} \,,
\label{e0}
\ee
and $c={1\over 2}$.  Standard computations give the partition functions
\be
\tr_{NS} e^{-\R H^{open}} &=& q^{-{1\over 48}} 
\prod_{n=0}^{\infty} \left( 1 + q^{{1\over 2}+n} \right) \,, \non \\
\tr_{NS} (-1)^{F} e^{-\R H^{open}} &=& q^{-{1\over 48}} 
\prod_{n=0}^{\infty} \left( 1 - q^{{1\over 2}+n} \right) \,, \non \\
\tr_{R} e^{-\R H^{open}} &=& 2 q^{{1\over 24}} 
\prod_{n=1}^{\infty} \left( 1 + q^{n} \right) \,, \non \\
\tr_{R} (-1)^{F} e^{-\R H^{open}} &=& 0 \,,
\ee
where $q=e^{- \pi \R/\LL}$ and $F$ is the
Fermion number operator.  For the case of the IM, the 
Virasoro algebra has three irreducible representations with
highest weights $0\,, 1/2 \,, 1/16$; the corresponding characters
(\ref{characters}) are given by (see, e.g., \cite{CFT})
\be
\chi_{0}(q) &=& {1\over 2} q^{-{1\over 48}} \left(
\prod_{n=0}^{\infty} \left( 1 + q^{{1\over 2}+n} \right) 
+ \prod_{n=0}^{\infty} \left( 1 - q^{{1\over 2}+n} \right)
\right) \,, \non \\
\chi_{1\over 2}(q) &=& {1\over 2} q^{-{1\over 48}} \left(
\prod_{n=0}^{\infty} \left( 1 + q^{{1\over 2}+n} \right) 
- \prod_{n=0}^{\infty} \left( 1 - q^{{1\over 2}+n} \right)
\right) \,, \non \\
\chi_{1\over 16}(q) &=&  q^{{1\over 24}} 
\prod_{n=1}^{\infty} \left( 1 + q^{n} \right) \,,
\ee
The partition functions therefore have the following expressions
in terms of characters
\be
\tr_{NS} e^{-\R H^{open}} &=& 
\chi_{0}(q) + \chi_{1\over 2}(q)\,, \non \\
\tr_{NS} (-1)^{F} e^{-\R H^{open}} &=&  
\chi_{0}(q) - \chi_{1\over 2}(q)\,, \non \\
\tr_{R} e^{-\R H^{open}} &=&  2 \chi_{1\over 16}(q)\,. 
\label{pfresults}
\ee
The modular transformation law (\ref{modtransf}) for the characters,
together with the explicit modular $S$ matrix for the case of the
IM (see, e.g. \cite{Ca2}), imply
\be
\chi_{0}(q) + \chi_{1\over 2}(q) &=& \chi_{0}(\tilde q) 
+ \chi_{1\over 2}(\tilde q)\,, \non \\
\chi_{0}(q) - \chi_{1\over 2}(q) &=& 
\sqrt{2}  \chi_{1\over 16}(\tilde q) \,, \non \\
\chi_{1\over 16}(q) &=&  {1\over \sqrt{2}}\left( \chi_{0}(\tilde q) 
- \chi_{1\over 2}(\tilde q)\right) \,,
\ee
where $\tilde q= e^{- 4 \pi \LL/\R}$. We conclude that the partition 
functions are given by
\be
\tr_{NS} e^{-\R H^{open}} &=& 
\chi_{0}(\tilde q) + \chi_{1\over 2}(\tilde q)\,, \non \\
\tr_{NS} (-1)^{F} e^{-\R H^{open}} &=&  
\sqrt{2}  \chi_{1\over 16}(\tilde q) \,, \non \\
\tr_{R} e^{-\R H^{open}} &=&  \sqrt{2} \left( \chi_{0}(\tilde q) 
- \chi_{1\over 2}(\tilde q)\right) \,, \non \\
\tr_{R} (-1)^{F} e^{-\R H^{open}} &=& 0 \,. 
\label{openchanresults}
\ee

\subsection{Closed channel}\label{subsec:closed}

In the closed channel, we regard $x$ as the time coordinate and 
$y$ as the space coordinate. Since $y$ is periodic, the fields $\psi$ 
and $\bar \psi$ can be either periodic ($R$) or anti-periodic ($NS$). These 
fields have the standard mode expansions
\be
\psi(x \,, y) &=& {1\over \sqrt{\R}}\sum_{k} a_{k}
e^{-i{2\pi\over \R} k (y - ix)} \,, \quad 
\left\{ a_{k} \,, a_{l} \right\} = \delta_{k+l\,, 0} \,, \non \\
\bar \psi(x \,, y) &=& {1\over \sqrt{\R}}\sum_{k} \bar a_{k}
e^{-i{2\pi\over \R} k (-y - ix)} \,, \quad 
\left\{ \bar a_{k} \,, \bar a_{l} \right\} = \delta_{k+l\,, 0} \,, 
\quad \left\{ a_{k} \,, \bar a_{l} \right\} = 0 \,,
\ee
with $k \in \Z$ for $R$, and $k \in \Z + {1\over 2}$ for $NS$. 
There are two independent sets of modes in the closed channel.

The Hamiltonian $H^{closed}$ is
\be
H^{closed} = {2\pi\over \R}\left(2e_{0} 
+ \sum_{k >0} k \left( a_{-k} a_{k} + \bar a_{-k} \bar a_{k} \right)
\right) = {2\pi\over \R}\left( L_{0} + \bar L_{0} - {c\over 12} \right) \,,
\ee
where again $k \in \Z$ for $R$, $k \in \Z + {1\over 2}$ for $NS$, and
$e_{0}$ is given in (\ref{e0}).

The boundary conditions (\ref{BC1}), (\ref{BC2}) now correspond to
initial and final conditions on states.  Expressing these conditions
in terms of modes, we are led to define (up to normalization) the
boundary kets $|B_{\pm} \rangle$ and the corresponding bras $\langle
B_{\pm}  |$ as the solutions of the constraints \cite{string}
\be
\left( a_{k} - i \gamma \bar a_{-k} \right) |B_{\gamma}  \rangle = 
0 \,, \qquad 
\langle B_{\gamma} | \left( a_{-k} + i \gamma \bar a_{k} \right) = 0
\,,
\label{IMtreeconstraint}
\ee
where $\gamma = \pm 1$. The solutions in the $NS$ sector are given by
\be
|B^{NS}_{\gamma}  \rangle = e^{i \gamma \sum_{k={1\over 2}}^{\infty} 
a_{-k} \bar a_{-k}} |0 \rangle \,, \qquad 
\langle B^{NS}_{\gamma}  | = \langle 0 | 
e^{-i \gamma \sum_{k={1\over 2}}^{\infty} \bar a_{k} a_{k}} \,,
\label{bsNS}
\ee
where the $NS$ vacuum state $|0 \rangle $ satisfies $a_{-k} |0 \rangle 
=0 \,, \quad \bar a_{-k} |0 \rangle =0$ for $k > 0$. 
These states have even Fermion parity
\be
(-1)^{F} |B^{NS}_{\gamma}  \rangle = |B^{NS}_{\gamma}  \rangle \,,
\ee 
where $F$ is now the total Fermion number operator in the $NS$ sector,
\be
F = \sum_{k={1\over 2}}^{\infty} \left( a_{-k} a_{k} 
+ \bar a_{-k} \bar a_{k} \right) \,.
\ee 

The solutions of (\ref{IMtreeconstraint}) in the $R$ sector are given by
\be
|B^{R}_{\gamma}  \rangle = e^{i \gamma \sum_{k=1}^{\infty} 
a_{-k} \bar a_{-k}} |\gamma \rangle \,, \qquad
\langle B^{R}_{\gamma}  | = \langle \gamma | 
e^{-i \gamma \sum_{k=1}^{\infty} \bar a_{k} a_{k}} \,,
\label{bsR}
\ee
where the degenerate $R$ vacuum states $|\pm \rangle $ satisfy $a_{-k}
|\pm \rangle =0 \,, \quad \bar a_{-k} |\pm \rangle =0$ for $k > 0$, as
well as
\be
\left( a_{0} - i \gamma \bar a_{0} \right)  |\gamma \rangle = 0 \,.
\label{bsRzero}
\ee
An explicit representation for the zero modes is (see, e.g., 
\cite{CFT, Ya})
\be
a_{0} |\pm \rangle &=& {1\over \sqrt{2}}e^{\pm i {\pi\over 4}} |\mp \rangle
\,, \non \\
\bar a_{0} |\pm \rangle &=& {1\over \sqrt{2}}e^{\mp i {\pi\over 4}} |\mp \rangle
\,,
\ee
using which one can readily verify (\ref{bsRzero}).
Moreover, $\left( 2 i a_{0} \bar a_{0} \right) |\pm \rangle = \pm |\pm \rangle$.
The total Fermion parity operator $(-1)^{F}$ in the $R$ sector is 
given by
\be
(-1)^{F} = \left( 2 i a_{0} \bar a_{0} \right) 
e^{i \pi  \sum_{k=1}^{\infty} \left( a_{-k} a_{k} 
+ \bar a_{-k} \bar a_{k} \right) } \,,
\ee
(we choose the sign so that $|+ \rangle$ has $(-1)^{F} = 1$), and thus,
the boundary states satisfy
\be
(-1)^{F} |B^{R}_{\pm}  \rangle = \pm  |B^{R}_{\pm} \rangle \,.
\label{RparIsing}
\ee

Using standard techniques, we find
\be
\langle B^{NS}_{\pm} | e^{-\LL H^{closed}} |B^{NS}_{\pm} \rangle
&=& \tilde q^{-{1\over 48}} 
\prod_{n=0}^{\infty} \left( 1 + \tilde q^{{1\over 2}+n} \right) 
= \chi_{0}(\tilde q) + \chi_{1\over 2}(\tilde q) \,, \non  \\
\langle B^{NS}_{\mp} | e^{-\LL H^{closed}} |B^{NS}_{\pm} \rangle
&=& \tilde q^{-{1\over 48}} 
\prod_{n=0}^{\infty} \left( 1 - \tilde q^{{1\over 2}+n} \right) 
= \chi_{0}(\tilde q) - \chi_{1\over 2}(\tilde q) \,, \non  \\
\langle B^{R}_{\pm} | e^{-\LL H^{closed}} |B^{R}_{\pm} \rangle
&=& \tilde q^{{1\over 24}} 
\prod_{n=1}^{\infty} \left( 1 + \tilde q^{n} \right) 
= \chi_{1\over 16}(\tilde q) \,, \non  \\
\langle B^{R}_{\mp} | e^{-\LL H^{closed}} |B^{R}_{\pm} \rangle
&=& 0 \,.
\label{closedchanresults}
\ee 
Recalling the results (\ref{openchanresults}) from the
open channel, we obtain the sought-after relations
\be
\tr_{NS} e^{-\R H^{open}} &=& 
\langle B^{NS}_{\pm} | e^{-\LL H^{closed}} |B^{NS}_{\pm} \rangle
\,, \non \\
\tr_{NS} (-1)^{F} e^{-\R H^{open}} &=& \sqrt{2} 
\langle B^{R}_{\pm} | e^{-\LL H^{closed}} |B^{R}_{\pm} \rangle
\,, \non \\
\tr_{R} e^{-\R H^{open}} &=&  \sqrt{2} 
\langle B^{NS}_{\mp} | e^{-\LL H^{closed}} |B^{NS}_{\pm} \rangle
\,, \non \\
\tr_{R} (-1)^{F} e^{-\R H^{open}} &=& 0 = 
\langle B^{R}_{\mp} | e^{-\LL H^{closed}} |B^{R}_{\pm} \rangle
\,, 
\label{relations}
\ee
which show explicitly how the $NS$ and $R$ sectors
``transform'' between the open and closed channels of the
cylinder.  \footnote{Numerical factors appear in these relations
because the $NS$ and $R$ sectors are not irreducible representations of
the Virasoro algebra and also the states $|B_{\pm}^{NS}\rangle$
are not properly normalized.  See Eqs. 
(\ref{IMIs}) and (\ref{normalization}) below.}
Similar results are known in string theory.

We conclude this subsection by noting that the boundary states
(\ref{bsNS}), (\ref{bsR}) are closely related to the Ishibashi states
(\ref{Ishibashi}). Namely,
\be
| 0 \rangle\rangle &=& {1\over 2}\left( |B^{NS}_{+}\rangle +
|B^{NS}_{-} \rangle \right) \,, \non \\
| {1\over 2} \rangle\rangle &=& {1\over 2}\left( |B^{NS}_{+} \rangle -
|B^{NS}_{-} \rangle \right) \,, \non \\
| {1\over 16} \rangle\rangle &=& |B^{R}_{+} \rangle  \,.
\label{IMIs}
\ee 
Indeed, recalling that the boundary states satisfy (\ref{IMtreeconstraint})
and that 
\be
L_{n} = {1\over 2}\sum_{k} (k + {n\over 2}) :a_{-k} a_{n+k}: \,, \qquad 
\bar L_{n} = {1\over 2}\sum_{k} (k + {n\over 2}) :\bar a_{-k} \bar a_{n+k}: \,,
\ee
one can easily show that the boundary states satisfy the 
constraint (\ref{treeconstraint}). Moreover, expanding the exponentials in 
the expressions (\ref{bsNS}), (\ref{bsR}) and comparing the leading 
terms with (\ref{Ishibashi}), one can infer (\ref{IMIs}).
Regarding the Ishibashi states as orthonormal vectors $( i \,, j ) =
\delta_{ij}$, it follows from (\ref{IMIs}) that the boundary states
have the normalization
\be
\left( B^{NS}_{\pm} \,, B^{NS}_{\pm} \right) = 2 \,, \qquad
\left( B^{R}_{\pm} \,, B^{R}_{\pm} \right) = 1 \,.
\label{normalization}
\ee 
Strictly speaking, the Ishibashi states
and boundary states $|B_{\pm} \rangle $ are {\it not} normalizable. However, 
one can define an inner product \cite{Is, Ca2} and argue
\be
{\left( B^{NS}_{\pm} \,, B^{NS}_{\pm} \right)\over
\left( 0 \,, 0 \right)} = \lim_{q \rightarrow 1} 
{\langle B^{NS}_{\pm} | q^{L_{0} + \bar L_{0}} |B^{NS}_{\pm} 
\rangle\over
\langle \langle 0 | q^{L_{0} + \bar L_{0}} | 0 \rangle \rangle}
= \lim_{q \rightarrow 1} {\chi_{0}(q^{2}) + \chi_{1\over 2}(q^{2}) 
\over \chi_{0}(q^{2})} = 2 \,.  
\ee

\subsection{Consistent boundary states}\label{subsec:IMconsistent}

Finally, it is also instructive to rederive Cardy's results for the
consistent IM boundary states in our basis $|B_{\pm} \rangle $.
We begin by rewriting the fundamental consistency constraint
(\ref{cylinderconstraint}) as
\be
\tr_{NS}{1\over 2}(1 + (-1)^{F}) e^{-\R H_{\alpha \beta}^{open}} +
\tr_{NS}{1\over 2}(1 - (-1)^{F}) e^{-\R H_{\alpha \beta}^{open}} \non \\
+\tr_{R}{1\over 2}(1 + (-1)^{F}) e^{-\R H_{\alpha \beta}^{open}} 
= \langle \alpha | e^{-\LL H^{closed}} |\beta \rangle \,.
\label{IMcylinderconstraint}
\ee
From the results (\ref{pfresults}), it is evident that 
\be
\tr_{NS}{1\over 2}(1 + (-1)^{F}) e^{-\R H_{\alpha \beta}^{open}} &=&
N^{0}_{\alpha \beta} \chi_{0}(q) \,, \non \\
\tr_{NS}{1\over 2}(1 - (-1)^{F}) e^{-\R H_{\alpha \beta}^{open}} &=&
N^{{1\over 2}}_{\alpha \beta} \chi_{{1\over 2}}(q) \,, \non \\
\tr_{R}{1\over 2}(1 + (-1)^{F}) e^{-\R H_{\alpha \beta}^{open}} &=&
N^{{1\over 16}}_{\alpha \beta} \chi_{{1\over 16}}(q) \,,
\ee 
and so the LHS of (\ref{IMcylinderconstraint}) is indeed equal to
$\sum_{i} N_{\alpha \beta}^{i} \chi_{i}(q)$. In the RHS of 
(\ref{IMcylinderconstraint}), we expand the boundary 
states in the basis $|B_{\pm} \rangle $ using
\be
| \alpha \rangle = {1\over 2}\left(
| B^{NS}_{+} \rangle \langle B^{NS}_{+} | \alpha \rangle + 
| B^{NS}_{-} \rangle \langle B^{NS}_{-} | \alpha \rangle \right) +
| B^{R}_{+} \rangle \langle B^{R}_{+} | \alpha \rangle \,,
\label{IMbasis}
\ee
keeping in mind the normalization (\ref{normalization}).  Then, making
use also of the relations (\ref{relations}), we arrive at the Cardy
equations
\be
N^{0}_{\alpha \beta} &=& 
{1\over 4}\langle \alpha |B^{NS}_{+} \rangle 
\langle B^{NS}_{+}  |\beta \rangle
+ {1\over 4}\langle \alpha |B^{NS}_{-} \rangle 
\langle B^{NS}_{-}  |\beta \rangle
+ {1\over \sqrt{2}}\langle \alpha |B^{R}_{+} \rangle 
\langle B^{R}_{+}  |\beta \rangle \,, \non  \\
N^{{1\over 2}}_{\alpha \beta} &=& 
{1\over 4}\langle \alpha |B^{NS}_{+} \rangle 
\langle B^{NS}_{+}  |\beta \rangle
+ {1\over 4}\langle \alpha |B^{NS}_{-} \rangle 
\langle B^{NS}_{-}  |\beta \rangle
- {1\over \sqrt{2}}\langle \alpha |B^{R}_{+} \rangle 
\langle B^{R}_{+}  |\beta \rangle \,, \non  \\
N^{{1\over 16}}_{\alpha \beta} &=& 
{1\over 2\sqrt{2}}\langle \alpha |B^{NS}_{+} \rangle 
\langle B^{NS}_{-}  |\beta \rangle
+ {1\over 2\sqrt{2}}\langle \alpha |B^{NS}_{-} \rangle 
\langle B^{NS}_{+}  |\beta \rangle \,.
\label{IMCa}
\ee 
Following Cardy \cite{Ca2}, we define the states $|{\bf k} \rangle$ by 
$N^{i}_{{\bf 0} {\bf k} }=\delta^{i}_{k}$, and we obtain
\be
|{\bf 0}\rangle  &=& {1\over \sqrt{2}}|B^{NS}_{-} \rangle 
+ {1\over \sqrt[4]{2}}|B^{R}_{+} \rangle  \,, \non  \\
|{\bf {1\over 2}}  \rangle  &=& {1\over \sqrt{2}}|B^{NS}_{-} \rangle 
- {1\over \sqrt[4]{2}}|B^{R}_{+} \rangle  \,, \non  \\
|{\bf {1\over 16}}  \rangle  &=&  |B^{NS}_{+} \rangle \,.
\label{IMstates}
\ee
These states correspond to the boundary conditions ``fixed $+$'', 
``fixed $-$'', and ``free'', respectively.
The $g$-factor \cite{AL} of a boundary state $| \alpha \rangle$ is 
given by 
\be
g_{\alpha} = \langle \langle 0 | \alpha \rangle =
{1\over 2}\left( \langle B^{NS}_{+}| + \langle B^{NS}_{-}|  \right)
| \alpha \rangle \,.
\ee
We therefore obtain (again remembering the normalization
(\ref{normalization})) the well-known results
\be
g_{{\bf 0}} = g_{{\bf {1\over 2}} } = {1\over \sqrt{2}} \,, \qquad
g_{{\bf {1\over 16}} } = 1 \,.
\ee

\end{document}